\documentclass[ALICE,manyauthors]{cernphprep}

\usepackage{rotating}

\begin{document}
%
%
\begin{titlepage}
\PHnumber{2012-012}                  
\PHdate{31 Jan 2012}            			 
%
%
\title{J/$\psi$ suppression at forward rapidity \\ in Pb-Pb collisions at $\sqrt{s_{\mathrm{NN}} } = 2.76$ TeV}
\ShortTitle{J/$\psi$ suppression at forward rapidity in Pb-Pb collisions at $\sqrt{s_{\mathrm{NN}} } = 2.76$ TeV}   
%

\Collaboration{ALICE Collaboration%
         \thanks{See Appendix~\ref{app:collab} for the list of collaboration  members}}
\ShortAuthor{ALICE Collaboration}      
\begin{abstract} 

The ALICE experiment has measured the inclusive J/$\psi$ production  in Pb-Pb collisions at $\sqrt{s_{\mathrm{NN}} } = 2.76$ TeV 
 down to zero transverse momentum  in the rapidity range $2.5 < y < 4$.  
A suppression of the inclusive J/$\psi$ yield in Pb-Pb is observed with respect to the one measured in pp collisions
scaled by the number of binary nucleon-nucleon collisions. 
The nuclear modification factor, integrated over the 0\%--80\% most central collisions, is 
$0.545 \pm 0.032 \rm{(stat.)} \pm 0.083 \rm{(syst.)}$ and does not
exhibit a significant dependence on the collision centrality. 
These features appear significantly different from measurements at lower collision energies.    
Models including J/$\psi$ production from charm quarks in a deconfined partonic phase can describe our data.
\end{abstract}

\end{titlepage}
%
\setcounter{page}{2}


Ultra-relativistic collisions of heavy nuclei aim at producing nuclear matter at high temperature and pressure. 
Under such conditions Quantum Chromodynamics predicts the existence of a deconfined state of  partonic matter, the quark-gluon plasma (QGP).
Among the possible probes of the QGP, heavy quarks are of particular interest since they are expected 
to be produced in the primary partonic scatterings and to coexist with the surrounding medium. 
Therefore, the measurement of  quarkonium states and hadrons with open heavy flavor is expected 
to provide essential information on the properties of the strongly-interacting system formed in the early stages of 
heavy-ion collisions~\cite{Bedjidian:2004gd}. In particular, according to the color-screening 
model~\cite{Matsui:1986dk}, measuring the in-medium dissociation probability of the different quarkonium states is expected to provide an 
estimate of the initial temperature of the system. 
In the past two decades, J/$\psi$ production in heavy-ion collisions was intensively studied at 
the Super Proton Synchrotron (SPS) and at the Relativistic Heavy Ion Collider (RHIC),
from  approximately  $20 $ to $200$~GeV center of mass energy per nucleon pair ($\sqrt{s_{\mathrm{NN}}}$). 
At the SPS, a strong J/$\psi$ suppression was found in the most central Pb-Pb collisions~\cite{Alessandro:2004ap}. 
The observed suppression is larger than the one expected from Cold Nuclear Matter (CNM) effects, which include nuclear absorption and 
(anti-) shadowing.
The dissociation of  excited $\rm{c \overline c}$ states like $\chi_{\rm{c}}$ and $\psi\rm{(2S)}$,  which in pp collisions 
constitute about 40\% of the inclusive J/$\psi$ yield~\cite{Bedjidian:2004gd}, is one possible interpretation of the observed suppression. 
A J/$\psi$ suppression was also observed at RHIC, in central Au-Au collisions~\cite{Adare:2006ns,Adare:2011yf}, 
at a level similar to the one observed at the SPS when measured at mid-rapidity although it is larger at forward rapidity. 
Several models~\cite{{BraunMunzinger:2000px},{Thews:2000rj},{Andronic:2007bi},{Zhao:2007hh}}
attempt to reproduce the RHIC data by adding to the direct J/$\psi$ production a regeneration component from deconfined charm quarks in the medium,  
which counteracts the J/$\psi$  dissociation in a QGP. 
A  quantitative description of these final-state effects is however difficult at the present time 
because of the lack of precision in the CNM effects and in the open charm cross section determination.
The measurement of charmonium production is especially promising at the Large Hadron Collider (LHC)
where the high-energy density of the medium  and the large number of  c$\bar{\rm{c}}$ pairs  
produced in central Pb-Pb collisions should help to disentangle between the different suppression and regeneration scenarios. 
At the LHC, a suppression of inclusive J/$\psi$ with high transverse momentum  was observed in central Pb-Pb  collisions
 at $\sqrt{s_{\mathrm{NN}}} = 2.76$ TeV  with respect to peripheral collisions or pp collisions at the same energy
 by ATLAS~\cite{Atlas:2010px} and CMS~\cite{Chatrchyan:2012np}, respectively.

In this Letter, we report ALICE results on inclusive  J/$\psi$ production in Pb-Pb collisions at $\sqrt{s_{\mathrm{NN}}} = 2.76$ TeV
at forward rapidity,  measured via the $\mu^{+} \mu^{-}$ decay channel.  
Our measurement encloses the low transverse momentum region that is not accessible to other LHC experiments and thus complements their observations.  
The J/$\psi$ corrected yield in Pb-Pb collisions is combined with the one measured in pp collisions  at the same center-of-mass energy~\cite{Aamodt:2011tmp} 
to form the J/$\psi$ nuclear modification factor $R_{\rm{AA}}$. 
The results are presented as a function of collision centrality and rapidity ($y$), and in intervals of transverse momentum ($p_{\rm{t}}$).

The ALICE detector is described in~\cite{Aamodt:2008zz}. 
At forward rapidity  ($2.5 < y < 4$)
the production of quarkonium states is measured in the muon spectrometer~\footnote{In the ALICE reference frame, 
the muon spectrometer covers a negative $\eta$ range and consequently a negative $y$ range.
We have chosen to present our results with a positive $y$ notation.}
down  to $p_{\rm{t}} = 0$. 
The spectrometer consists of a ten interaction length thick absorber filtering the 
muons in front of five tracking stations comprising two planes of cathode pad chambers each, with the third station 
inside a dipole magnet with  a 3 Tm field integral. 
The tracking apparatus is completed by a triggering system made of four planes of resistive plate chambers
 downstream of a 1.2 m thick iron wall,  
which absorbs secondary hadrons escaping from the front absorber and low momentum  muons coming mainly from $\pi$ and K decays.
In addition, the silicon pixel detector (SPD) and scintillator arrays (VZERO) were  used in this analysis.  
The VZERO counters, two arrays of 32 scintillator tiles each, cover  
 $2.8 \leq \eta \leq 5.1$ (VZERO-A) and $-3.7 \leq \eta \leq -1.7$ (VZERO-C).
The SPD consists of two cylindrical layers 
covering  $|\eta|  \leq 2.0$  and $|\eta| \leq 1.4$  for the inner and outer layers, respectively. 
All these detectors have full azimuthal coverage.
The minimum bias (MB) trigger requirement used for this analysis 
consists of a logical \textsc{AND}  of the three following conditions: 
(i) a signal in  two  readout chips in the outer layer of the SPD, 
(ii) a signal in VZERO-A, 
(iii) a signal in VZERO-C, 
 providing a high triggering efficiency for hadronic interactions.
The beam induced  background was further reduced by timing cuts on the signals from  the VZERO  and from the zero degree 
calorimeters (ZDC). 
The contribution from electromagnetic processes was removed  with a cut on the energy deposited in the neutron ZDCs.
The centrality determination is based on a fit of the VZERO amplitude distribution as described  in~\cite{PhysRevLett.106.032301}.
A cut corresponding to the most central 80\%  of the nuclear cross section was applied; 
for these events the MB trigger is fully efficient.
A data sample of 17.7 $\times$ 10$^{6}$ Pb-Pb collisions collected in 2010 satisfying all the above conditions
 is used in the following analysis. It corresponds to an integrated luminosity 
$\mathcal{L}_{\rm int} \approx 2.9 $  $\mu$b$^{-1}$. 
This data sample was further divided into five centrality classes from 0\%--10\% (central collisions) to 
50\%--80\% (peripheral collisions).

J/$\psi$ candidates are formed by combining pairs of opposite-sign (OS) tracks reconstructed in 
the geometrical acceptance of the muon spectrometer.
\begin{figure}
\includegraphics[width=0.99\linewidth,keepaspectratio]{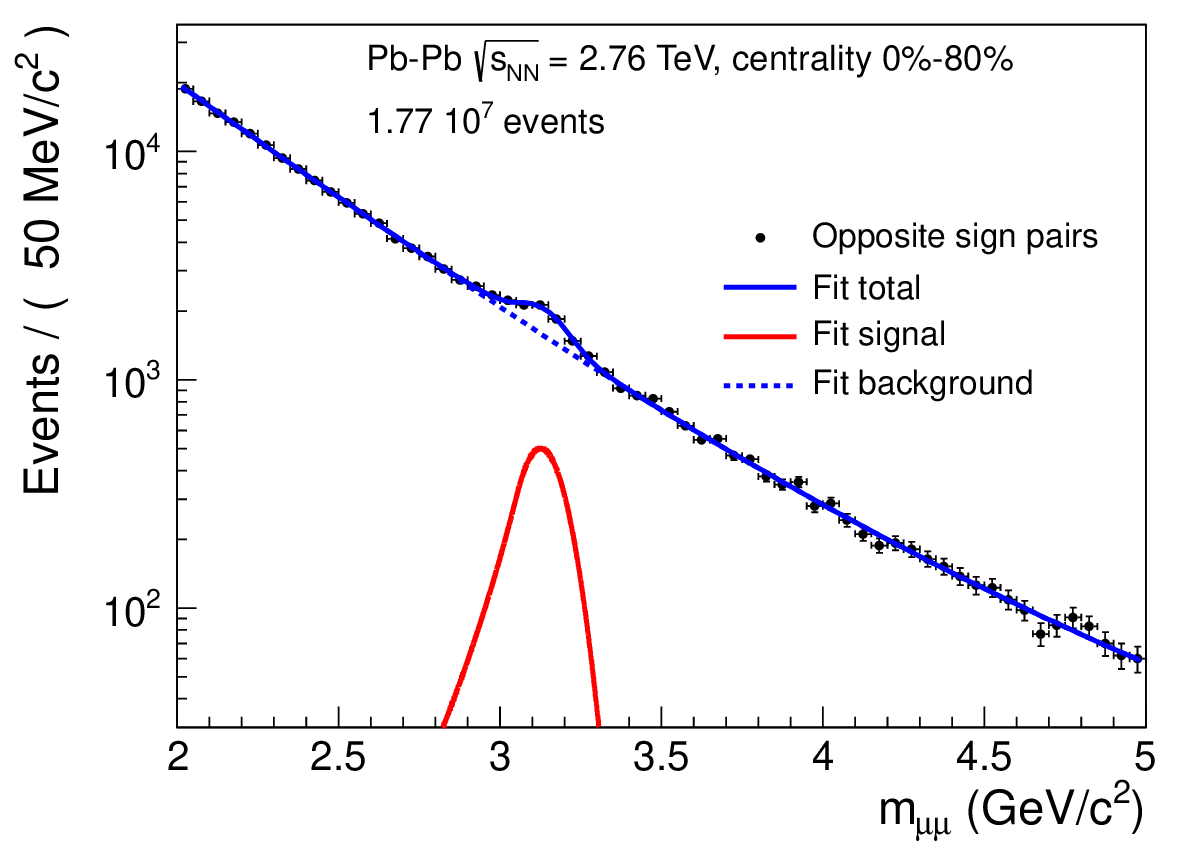}
\caption{(Color online) \label{fig:invmassfit} Invariant mass spectrum of $\mu^{+} \mu^{-}$  pairs (solid black circles)
with $p_{\rm{t}} \geq 0$  and $2.5 < y < 4 $  in the 0\%--80\%  most central Pb-Pb collisions.}
\end{figure}
To reduce the combinatorial background, the reconstructed tracks in the muon tracking chambers
 are required to match a track segment  in the muon trigger system.
The resulting invariant mass distribution of OS  muon pairs for the 0\%--80\%  most central Pb-Pb collisions is shown in Fig.~\ref{fig:invmassfit}, where
a J/$\psi$ signal above the combinatorial background is clearly visible.  
The J/$\psi$ raw yield was extracted by using two different methods. 
The OS dimuon invariant mass distribution was fitted with a Crystal Ball (CB) function to reproduce the J/$\psi$ line shape, and a sum of two exponentials to describe the
underlying continuum. The CB function connects a Gaussian core with a power-law tail~\cite{CBdef} at low mass to account for energy loss fluctuations and radiative decays.
At high transverse momenta ($p_{\rm{t}} \geq 3 \ {\rm GeV}/c$), the sum of two exponentials does not
 describe correctly  the underlying continuum; it was replaced by a third order polynomial. 
Alternatively,  the combinatorial background was subtracted using an event-mixing technique. 
The resulting mass distribution was fitted with a CB function and an exponential or a first order polynomial
 to describe the remaining background.
The event mixing was preferred to the like-sign subtraction technique since it is less sensitive to correlated signal pairs 
present in the like-sign spectra and gives better statistical precision.
The $\psi \rm{(2S)}$ was not included in the signal line shape since its contribution is negligible.
The width of the J/$\psi$ mass peak depends on the resolution of the spectrometer
which can be affected by the detector occupancy that increases with centrality. 
This effect, evaluated by embedding simulated J/$\psi \rightarrow \mu^{+} \mu^{-}$ decays into real events,  
was found to be less than 2\%. 
This conclusion was confirmed by a direct measurement of the tracking chamber resolution versus centrality using reconstructed tracks. 
Therefore, the same CB line shape can be used for all centrality classes. 
The parameters of the CB tails were fixed to the values obtained  either in simulations 
or in pp collisions where the signal to background ratio is much higher. 
For each of these choices, the mean and width of the CB Gaussian part were fixed 
to the value obtained by fitting  the mass distribution in the centrality range 0\%--80\%.
In addition, a variation of the  width of $\pm$~1 standard deviation was applied to
account for uncertainties (varying the mean has turned out to have a negligible effect in comparison). 
The raw J/$\psi$ yield in each centrality class was determined as the average of the 
results obtained with the two fitting approaches and the various CB parametrizations, while the 
corresponding systematic uncertainties were defined as the RMS of these results. 
It was also checked that every individual result differs from the mean value by less than three RMS. 
The raw J/$\psi$ yield in our Pb-Pb sample is $2350 \pm  139 \rm{(stat.)} \pm 189 \rm{(syst.)}$. 
The invariant mass resolution is around 78~MeV/$c^{2}$, in very good agreement with the embedded J/$\psi$ simulations.
The signal to background ratio integrated over $\pm \  3 \  \sigma$ of the mass resolution varies from  0.1 for central collisions to 1.5 for peripheral collisions.

The measured number of J/$\psi$ ($N_{\rm{J/}\psi}^{i}$) was normalized 
to the number of events in the corresponding centrality class ($N_{\rm{events}}^{i}$) and further corrected for the 
branching ratio (BR) of the dimuon decay channel, the acceptance $A$ and the efficiency $\epsilon^{i}$  of the detector. 
The  inclusive J/$\psi$ yield in each centrality class $i$  for our measured $p_{\rm{t}}$ and $y$ ranges ($\Delta  p_{\rm{t}}$, $\Delta y$) is then given by:
\begin{eqnarray}
Y_{\rm{J/}\psi}^{i} (\Delta  p_{\rm{t}}, \Delta y )  =  \frac{N_{\rm{J/}\psi}^{i}}{{\rm BR}_{\rm{J/}\psi\rightarrow\mu^{+}\mu^{-}} N_{\rm{events}}^{i}  A \epsilon^{i} }.
\label{eq:jpsiyield}
\end{eqnarray}
The product $A\epsilon$  was determined from Monte Carlo simulations.
The generated J/$\psi$ $p_{\rm{t}}$ and $y$ distributions were extrapolated from existing
measurements~\cite{Bossu:2011qe}, including shadowing effects from EKS98 calculations~\cite{eskola-1999-9}.
As the measured J/$\psi$ polarization in pp collisions at $\sqrt{s} = 7$ TeV is compatible with zero~\cite{Abelev:2011md}, and J/$\psi$ mesons
  produced from charm quarks in the medium are expected to be unpolarized, we presume J/$\psi$ production is unpolarized.
For the tracking chambers, the time-dependent status of each electronic channel during the data taking period was taken 
into account as well as the residual misalignment of the detection elements.
The efficiencies of the muon trigger chambers were determined from data and were then
applied in the simulations~\cite{aliceMuonTrigChEff:2008}.  
Finally, the dependence of the efficiency on the detector occupancy was included using the embedding technique. 
For J/$\psi$ mesons emitted at $2.5 < y < 4 $ and $p_{\rm{t}} \geq 0$, 
a run-averaged value of $A\epsilon = 0.176$ with a 8\% relative systematic uncertainty was obtained. 
A $ 8\%  \pm  2\% ({\rm syst.})$ relative decrease of the efficiency was observed when going from peripheral to central collisions.

The J/$\psi$  yield measured in Pb-Pb collisions in centrality class $i$ is combined with the inclusive J/$\psi$ cross section
measured in pp collisions at the same energy to form the  nuclear modification factor $R_{\rm{AA}}$ defined as:
\begin{eqnarray}
R_{\rm{AA}}^{i} = \frac{Y_{\rm{J/}\psi}^{i} (\Delta  p_{\rm{t}}, \Delta y ) }{\langle T_{\rm{AA}}^{i} \rangle \times \sigma_{\rm{J/}\psi}^{\rm{pp}} (\Delta  p_{\rm{t}}, \Delta y )}.
\label{eq:raa}
\end{eqnarray}
The inclusive J/$\psi$  cross section in pp collisions  $\sigma_{\rm{J/}\psi}^{\rm{pp}} (\Delta  p_{\rm{t}}, \Delta y )$ 
was measured using the same apparatus and analysis technique  within the corresponding $p_{\rm{t}}$ and  $y$ range~\cite{Aamodt:2011tmp}. 
The reference value $\sigma_{\rm{J/}\psi}^{\rm{pp}}$ used for the calculation of $R_{\rm{AA}}$ integrated over $p_{\rm{t}}$
and $y$ is  $3.34 \pm 0.13 \rm{(stat.)} \pm 0.24 \rm{(syst.)} \pm 0.12 \rm{(lumi.)} ^{+0.53}_{-1.07} \rm{(pol.)} \; \mu b$. 
The centrality intervals used in this analysis,  
the average number of participating nucleons $\langle N_{\rm{part}} \rangle$  and average value of the 
nuclear overlap function $\langle T_{\rm{AA}} \rangle$ derived from a Glauber model calculation~\cite{PhysRevLett.106.032301} 
are summarized in Table~\ref{tab:taa}.
\begin{table}
\caption{\label{tab:taa} The average number of participating nucleons $\langle N_{\rm{part}} \rangle$  without and with $\langle N_{\rm{coll}} \rangle$ weighting, 
 the mid-rapidity charged-particle density  d$N_{\rm{ch}}^{\rm{w}}/$d$\eta\vert_{\eta=0}$ with $\langle N_{\rm{coll}} \rangle$ weighting 
and the average value of the nuclear overlap function $\langle T_{\rm{AA}} \rangle$ 
for the  centrality classes  expressed in percentages of the nuclear cross section~\cite{PhysRevLett.106.032301}.}
\begin{center}
\begin{tabular}{cccccc}
Centrality   & $\langle N_{\rm{part}} \rangle$ &  $\langle N_{\rm{part}}^{\rm{w}} \rangle$  &  $\frac{ {\rm d} N_{\rm{ch}}^{\rm{w}} }{\rm{d}\eta\vert_{\eta=0}}$  &  $\langle T_{\rm{AA}}  \rangle$ (mb$^{-1}$)   \\
\hline
0\%--10\%     & 356$\pm$4     &   361$\pm$4    &     1463$\pm$60    &  23.5$\pm$1.0      \\
10\%--20\%    & 260$\pm$4     &   264$\pm$4    &     ~979$\pm$37    &  14.4$\pm$0.6      \\
20\%--30\%    & 186$\pm$4     &   189$\pm$4    &     ~658$\pm$23    &  8.74$\pm$0.37     \\
30\%--50\%    & 107$\pm$3     &   117$\pm$3    &     ~369$\pm$13    &  3.87$\pm$0.18     \\
50\%--80\%    & ~32$\pm$2     &   ~47$\pm$2    &     ~110$\pm$5     &  0.72$\pm$0.05     \\
0\%--80\%     & 139$\pm$3     &   264$\pm$4    &         --         &  7.03$\pm$0.27   
\end{tabular}
\end{center}
\end{table}
Since our most peripheral bin is rather large, the variables $\langle N_{\rm{part}}^{\rm{w}} \rangle$ and 
the charged-particle density measured at mid-rapidity ${\rm d}N_{\rm{ch}}^{\rm{w}}/{\rm d}\eta\vert_{\eta=0}$ were weighted   
by the number of binary collisions $\langle N_{\rm{coll}} \rangle$.
Indeed in absence of nuclear matter effects, the J/$\psi$ production cross section in nucleus-nucleus is expected to 
scale with $\langle N_{\rm{coll}} \rangle$.
The weighted values are given  in Table~\ref{tab:taa} and are used for the ALICE data points in the following figures. 
All systematic uncertainties entering the $R_{\rm{AA}}$ calculation are listed in Table~\ref{tab:syserr}. 
In the figures below, the point to point uncorrelated systematic uncertainties  are represented as boxes at 
the position of the data points  while the statistical ones are indicated by vertical bars. 
Correlated systematic uncertainties are quoted directly on the figures.
\begin{table}
\caption{\label{tab:syserr} Summary of the systematic uncertainties 
entering the  $R_{\rm{AA}}$ calculation. 
The type I (II) stands for correlated (uncorrelated) uncertainties. 
The centrality dependence  for the type II is given as a range.}	
\begin{center}
\begin{tabular}{lcc}
source                   				    & value   	        	& type  \\
\hline	
signal extraction						    & 5\%--12\%				& II    \\
input MC parametrization 					& 5\%             		& I     \\ 
tracking efficiency         			    & 5\% and 0\%--1\%	    & I and II \\
trigger efficiency						    & 4\% and 0\%--2\%		& I and II \\ 	
matching efficiency           				& 2\%					& I     \\ 
$T_{\rm{AA}}$								& 4\%--8\%				& II \\
$\sigma_{\rm{J/}\psi}^{\rm{pp}}$  at $\sqrt{s_{\mathrm{NN}}} = 2.76$ TeV   & 9\%            	& I \\   
\end{tabular}
\end{center}
\end{table}

\begin{figure*}
\begin{center}
\includegraphics[width=0.98\linewidth,keepaspectratio]{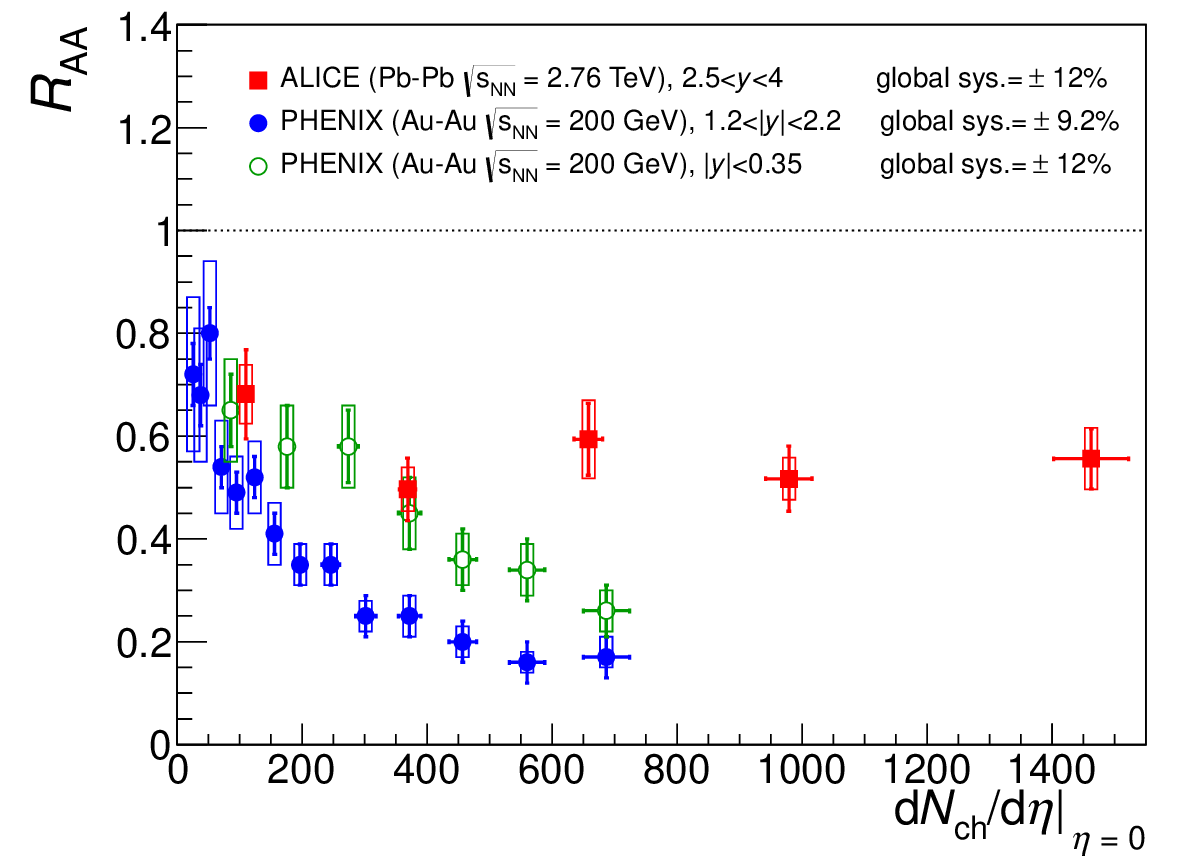}
\includegraphics[width=0.98\linewidth,keepaspectratio]{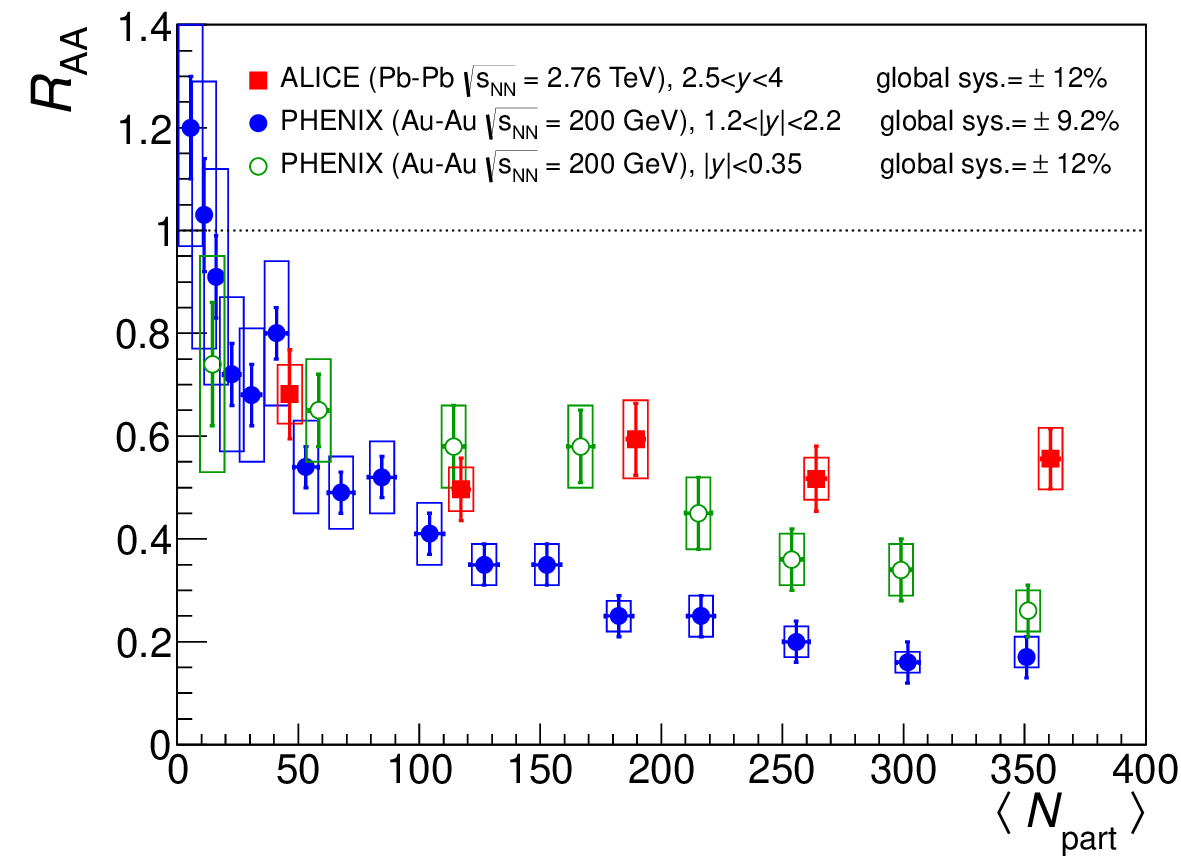}
\end{center}
\caption{(Color online) \label{fig:raa}  
Inclusive J/$\psi$ $R_{\rm{AA}}$ as a function of the mid-rapidity charged-particle density (top)  and the number of participating nucleons (bottom)
measured  in Pb-Pb collisions at $\sqrt{s_{\mathrm{NN}}} = 2.76$ TeV compared to PHENIX 
results in Au-Au collisions at $\sqrt{s_{\mathrm{NN}}} = 200$ GeV at 
mid-rapidity and forward rapidity~\cite{Adare:2006ns,Adare:2011yf,PhysRevC.71.049901}.
The ALICE data points are placed at the ${\rm d}N_{\rm{ch}}^{\rm{w}}/{\rm d}\eta\vert_{\eta=0}$  and  $\langle N_{\rm{part}}^{\rm{w}} \rangle$ values defined in Table~\ref{tab:taa}. }
\end{figure*}

The inclusive J/$\psi$ $R_{\rm{AA}}$ measured by ALICE at $\sqrt{s_{\mathrm{NN}}} = 2.76$ TeV in the range $2.5 < y < 4$ and $p_{\rm{t}} \geq 0$  
is shown in Fig.~\ref{fig:raa} as a function of 
$ {\rm d} N_{\rm{ch}}/{\rm d} \eta\vert_{\eta=0}$ (left) and $N_{\rm{part}}$ (right). The charged-particle density 
closely relates to the energy density of the created medium whereas the number of participants reflects the
collision geometry.
The  centrality integrated J/$\psi$ $R_{\rm{AA}}$ 
is $R^{0\%-80\%}_{\rm{AA}} = 0.545 \pm  0.032 \rm{(stat.)} \pm 0.083 \rm{(syst.)}$, indicating a clear J/$\psi$ suppression.
The contribution from  beauty hadron feed-down to the inclusive J/$\psi$ yield in our $y$ and $p_{\rm{t}}$ domain was measured 
by the LHCb collaboration to be about 10\%  in pp collisions at $\sqrt{s} = 7$ TeV~\cite{Aaij:2011jh}. 
Therefore, the difference between the prompt J/$\psi$  $R_{\rm{AA}}$
and our inclusive measurement is expected  not to exceed  11\% if  $N_{\rm{coll}}$ 
scaling of beauty production is assumed and shadowing effects are neglected.
All $R_{\rm{AA}}$ results are presented assuming unpolarized J/$\psi$ production in pp and Pb-Pb collisions.
The comparison with the PHENIX measurements~\footnote{
The PHENIX mid-rapidity J/$\psi$ $R_{\rm{AA}}$ was measured in centrality classes wider than the ones in which 
the mid-rapidity charged-particle density is given~\cite{PhysRevC.71.049901}.
Therefore  a linear interpolation was done to extract the mid-rapidity charged-particle density in the three most peripheral classes.}
 at $\sqrt{s_{\mathrm{NN}}} = 200$ GeV  at forward 
rapidity  $1.2 < |y| < 2.2$ \cite{Adare:2011yf,PhysRevC.71.049901}
shows that our inclusive J/$\psi$  $R_{\rm{AA}}$ is almost a factor of three larger for 
d$N_{\rm{ch}}/$d$\eta\vert_{\eta=0} \gtrsim 600$  ($N_{\rm{part}} \gtrsim 180$).  
In addition, our results do not exhibit a significant centrality dependence.

\begin{figure}
\begin{center}
\includegraphics[width=0.98\linewidth,keepaspectratio]{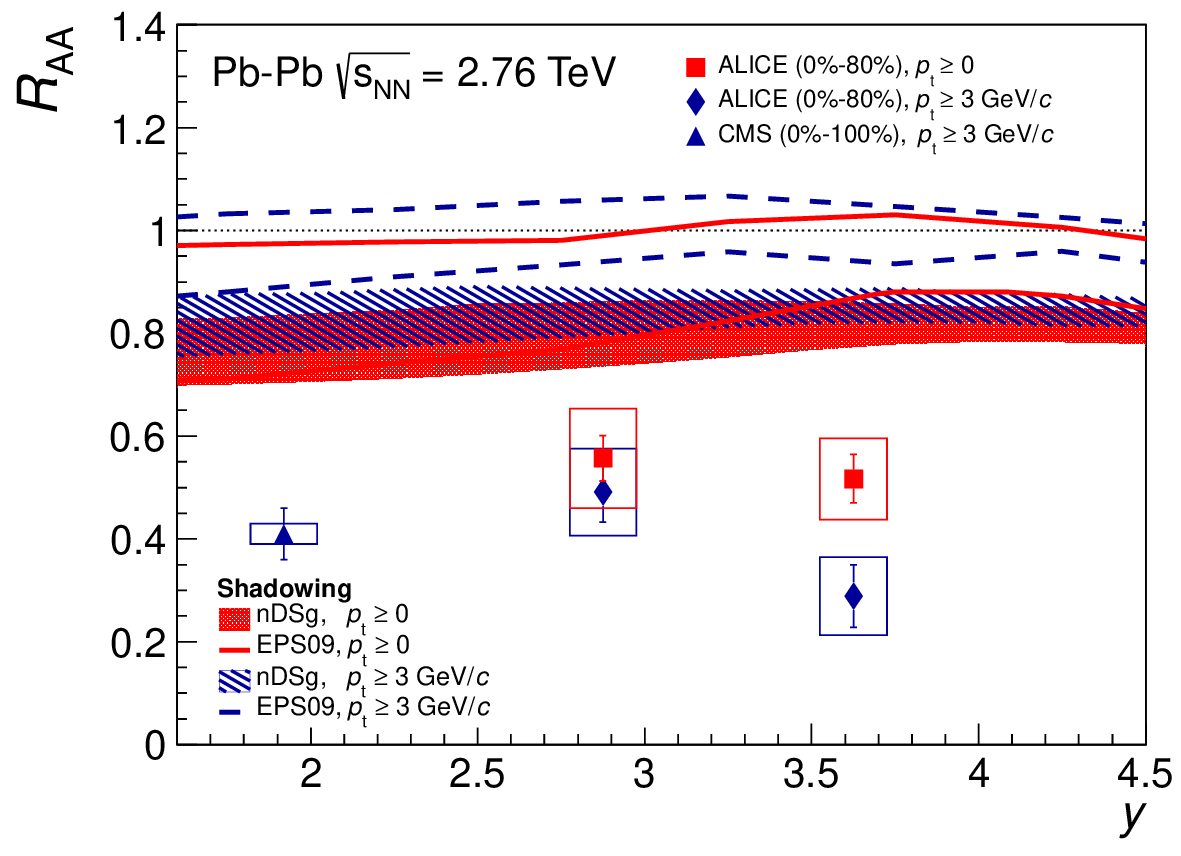}
\caption{(Color online) \label{fig:raavsy}  Centrality integrated inclusive J/$\psi$ $R_{\rm{AA}}$ measured in Pb-Pb 
collisions at $\sqrt{s_{\mathrm{NN}}} = 2.76$ TeV as a function of  rapidity  for two $p_{\rm{t}}$ ranges. 
The open boxes contain the total systematic uncertainties except the ones on the integrated luminosity in the pp reference 
 and on the $T_{\rm{AA}}$, i.e.  5.2\% (8.3\%) for the  ALICE (CMS~\cite{Chatrchyan:2012np}) data.
The two models~\cite{Ferreiro:2011rw,Vogt:2010aa} predict the  $R_{\rm{AA}}$  due  only to shadowing effects
for  nDSg (shaded areas) and EPS09 (lines) nPDF respectively.}
\end{center}
\end{figure}

The rapidity dependence of the J/$\psi$ $R_{\rm{AA}}$  is presented in Fig.~\ref{fig:raavsy}
for two $p_{\rm{t}}$ domains, $p_{\rm{t}} \geq$ 0 and  $p_{\rm{t}} \geq 3 \ {\rm GeV} /c$.  
The J/$\psi$ reference cross sections in pp collisions~\footnote{We report here 
$\sigma_{\rm{J/}\psi}^{\rm{pp}} (p_{\rm{t}} \geq 3 \rm{GeV}/c,\; 2.5 < y \leq 3.25) = 0.34 \pm 0.03 \rm{(stat.)} \pm 0.03 \rm{(syst.)} \pm 0.02 \rm{(lumi.)}  \; \mu b$ 
and 
$\sigma_{\rm{J/}\psi}^{\rm{pp}} (p_{\rm{t}} \geq 3 \rm{GeV}/c,\; 3.25 < y < 4)      = 0.50 \pm 0.04 \rm{(stat.)} \pm 0.04 \rm{(syst.)} \pm 0.02 \rm{(lumi.)} \; \mu b$
that can not directly be extracted from~\cite{Aamodt:2011tmp}.}  
and the $R_{\rm{AA}}$ total systematic uncertainties, 
indicated as open boxes in the figure, were evaluated in the same kinematic range. 
Our results are shown together with a measurement from CMS~\cite{Chatrchyan:2012np}
of the inclusive J/$\psi$ $R_{\rm{AA}}$  in the rapidity range $ 1.6 < |y| < 2.4 $ with $p_{\rm{t}} \geq 3 \ {\rm GeV} /c$. 
No significant rapidity dependence can be seen in the J/$\psi$ $R_{\rm{AA}}$  for $p_{\rm{t}} \geq 0$.   
For $p_{\rm{t}} \geq 3 \ {\rm GeV}/c$, a decrease of $R_{\rm{AA}}$  is observed with  increasing rapidity
reaching a value of $ 0.289 \pm  0.061 \rm{(stat.)} \pm 0.078 \rm{(syst.)}$ for $3.25 < y < 4$.
At LHC energies, J/$\psi$ nuclear absorption is likely to be negligible and the modification of the gluon distribution 
function is dominated  by shadowing effects~\cite{Lourenco:2008sk}.
An estimate of shadowing effects is shown in  Fig.~\ref{fig:raavsy} within the Color Singlet Model at Leading 
Order~\cite{Ferreiro:2011rw} and the Color Evaporation Model at Next to Leading Order~\cite{Vogt:2010aa}.
The shadowing is respectively calculated with the nDSg and the EPS09 parametrizations~\cite{Vogt:2010aa}  of the nuclear Parton Distribution Function (nPDF). 
For nDSg (EPS09) the upper and lower limits correspond to the uncertainty in the factorization scale (uncertainty of the nPDF).
The effect of shadowing shows no dependence with rapidity and its overall amount is reduced by the addition 
of a transverse momentum cut. At most, shadowing effects are expected to lower the $R_{\rm{AA}}$ from 1 to 0.7. 
Recent Color Glass Condensate (CGC) calculations for LHC energies
may indicate a larger initial state suppression ($R_{\rm{AA}} \approx 0.5$)~\cite{Dominguez:2011cy}.  
However, any J/$\psi$ suppression due to initial state effects, CGC or shadowing, will be stronger 
at lower $p_{\rm{t}}$ contrary to the data behavior.

\begin{figure}
\begin{center}
\includegraphics[width=0.98\linewidth,keepaspectratio]{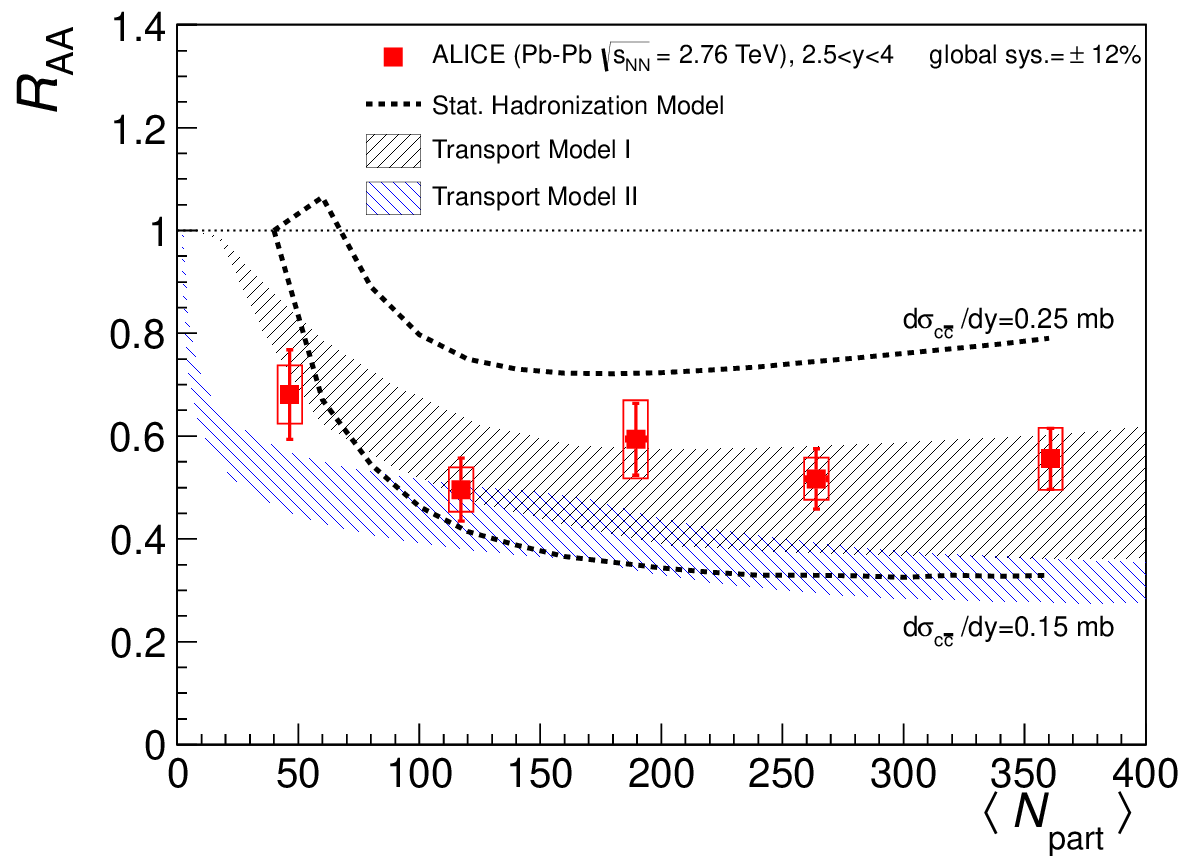}
\caption{(Color online) \label{fig:raamodels} Inclusive J/$\psi$ $R_{\rm{AA}}$ measured in Pb-Pb collisions at $\sqrt{s_{\mathrm{NN}}} = 2.76$ TeV compared to
 the predictions by  Statistical Hadronization Model~\cite{Andronic:2011yq}, 
 Transport Model I~\cite{Zhao:2011cv} and II~\cite{Liu:2009nb}, see text for details.
 The ALICE data points are placed at the $\langle N_{\rm{part}}^{\rm{w}} \rangle$ values defined in Table~\ref{tab:taa}. }
\end{center}
\end{figure}
In Fig.~\ref{fig:raamodels}, our measurement is compared with theoretical models that include 
a J/$\psi$ regeneration component from  deconfined charm quarks 
 in the medium. The Statistical Hadronization Model~\cite{BraunMunzinger:2000px,Andronic:2011yq} 
assumes deconfinement and a thermal equilibration of the bulk of the  c{$\bar{\rm{c}}$} pairs. 
Then charmonium production occurs only at phase boundary by statistical hadronization of
 charm quarks. The prediction is given for two values of 
$ {\rm d}\sigma_{\rm{c}\bar{\rm{c}}}/{\rm d}y$ in absence of a measurement for Pb-Pb collisions. 
The two transport  model results~\cite{Zhao:2011cv,Liu:2009nb}  presented in the same figure differ mostly in the rate 
equation controlling the J/$\psi$ dissociation and regeneration. 
Both are shown as a band which connects 
the results obtained with (lower limit) and without (higher limit) shadowing. 
The width of the band can be interpreted  as the uncertainty of the prediction.
In both transport models, the amount of regenerated J/$\psi$ in the most central 
collisions contributes to about 50\% of the production yield, the rest being from initial production.

In summary, we have presented the first  measurement of inclusive J/$\psi$ nuclear modification factor
down to $p_{\rm{t}} = 0$  at forward rapidity in Pb-Pb collisions at $\sqrt{s_{\mathrm{NN}}} = 2.76$ TeV. 
The J/$\psi$ $R_{\rm{AA}}$   is larger than the one measured at the SPS and at RHIC for most central collisions
and does not exhibit a significant  centrality dependence. 
Statistical hadronization and transport models which respectively feature a full and a partial J/$\psi$ production 
from charm quarks in the QGP phase can describe the data.
Towards a definitive conclusion about the role of J/$\psi$ 
production from deconfined charm quarks 
in a partonic phase, the amount of shadowing needs to be  measured  precisely in 
  pPb collisions. In this context, the measurement of open charm and J/$\psi$
elliptic flow will also help to determine the degree of thermalization for charm quarks. 

\newenvironment{acknowledgement}{\relax}{\relax}
\begin{acknowledgement}
\section*{Acknowledgements}
The ALICE collaboration would like to thank all its engineers and technicians for their invaluable contributions to the construction of the experiment and the CERN accelerator teams for the outstanding performance of the LHC complex.
\\
The ALICE collaboration acknowledges the following funding agencies for their support in building and
running the ALICE detector:
 \\
Calouste Gulbenkian Foundation from Lisbon and Swiss Fonds Kidagan, Armenia;
 \\
Conselho Nacional de Desenvolvimento Cient\'{\i}fico e Tecnol\'{o}gico (CNPq), Financiadora de Estudos e Projetos (FINEP),
Funda\c{c}\~{a}o de Amparo \`{a} Pesquisa do Estado de S\~{a}o Paulo (FAPESP);
 \\
National Natural Science Foundation of China (NSFC), the Chinese Ministry of Education (CMOE)
and the Ministry of Science and Technology of China (MSTC);
 \\
Ministry of Education and Youth of the Czech Republic;
 \\
Danish Natural Science Research Council, the Carlsberg Foundation and the Danish National Research Foundation;
 \\
The European Research Council under the European Community's Seventh Framework Programme;
 \\
Helsinki Institute of Physics and the Academy of Finland;
 \\
French CNRS-IN2P3, the `Region Pays de Loire', `Region Alsace', `Region Auvergne' and CEA, France;
 \\
German BMBF and the Helmholtz Association;
\\
General Secretariat for Research and Technology, Ministry of
Development, Greece;
\\
Hungarian OTKA and National Office for Research and Technology (NKTH);
 \\
Department of Atomic Energy and Department of Science and Technology of the Government of India;
 \\
Istituto Nazionale di Fisica Nucleare (INFN) of Italy;
 \\
MEXT Grant-in-Aid for Specially Promoted Research, Ja\-pan;
 \\
Joint Institute for Nuclear Research, Dubna;
 \\
National Research Foundation of Korea (NRF);
 \\
CONACYT, DGAPA, M\'{e}xico, ALFA-EC and the HELEN Program (High-Energy physics Latin-American--European Network);
 \\
Stichting voor Fundamenteel Onderzoek der Materie (FOM) and the Nederlandse Organisatie voor Wetenschappelijk Onderzoek (NWO), Netherlands;
 \\
Research Council of Norway (NFR);
 \\
Polish Ministry of Science and Higher Education;
 \\
National Authority for Scientific Research - NASR (Autoritatea Na\c{t}ional\u{a} pentru Cercetare \c{S}tiin\c{t}ific\u{a} - ANCS);
 \\
Federal Agency of Science of the Ministry of Education and Science of Russian Federation, International Science and
Technology Center, Russian Academy of Sciences, Russian Federal Agency of Atomic Energy, Russian Federal Agency for Science and Innovations and CERN-INTAS;
 \\
Ministry of Education of Slovakia;
 \\
Department of Science and Technology, South Africa;
 \\
CIEMAT, EELA, Ministerio de Educaci\'{o}n y Ciencia of Spain, Xunta de Galicia (Conseller\'{\i}a de Educaci\'{o}n),
CEA\-DEN, Cubaenerg\'{\i}a, Cuba, and IAEA (International Atomic Energy Agency);
 \\
Swedish Research Council (VR) and Knut $\&$ Alice Wallenberg
Foundation (KAW);
 \\
Ukraine Ministry of Education and Science;
 \\
United Kingdom Science and Technology Facilities Council (STFC);
 \\
The United States Department of Energy, the United States National
Science Foundation, the State of Texas, and the State of Ohio.
\end{acknowledgement}

\bibliographystyle{ieeetr}
\bibliography{./JpsiToMuonPbPbprep_v16.bib}

\newpage
%
%
\appendix
\section{The ALICE Collaboration}
\label{app:collab}

\begingroup
\small
\begin{flushleft}
B.~Abelev\Irefn{org1234}\And
J.~Adam\Irefn{org1274}\And
D.~Adamov\'{a}\Irefn{org1283}\And
A.M.~Adare\Irefn{org1260}\And
M.M.~Aggarwal\Irefn{org1157}\And
G.~Aglieri~Rinella\Irefn{org1192}\And
A.G.~Agocs\Irefn{org1143}\And
A.~Agostinelli\Irefn{org1132}\And
S.~Aguilar~Salazar\Irefn{org1247}\And
Z.~Ahammed\Irefn{org1225}
% ...
\renewcommand\labelenumi{\textsuperscript{\theenumi}~}
\section*{Affiliation notes}
\renewcommand\theenumi{\roman{enumi}}
\begin{Authlist}
\item \Adef{M.V.Lomonosov Moscow State University, D.V.Skobeltsyn Institute of Nuclear Physics, Moscow, Russia}Also at: M.V.Lomonosov Moscow State University, D.V.Skobeltsyn Institute of Nuclear Physics, Moscow, Russia
\item \Adef{Institute of Nuclear Sciences, Belgrade, Serbia}Also at: "Vin\v{c}a" Institute of Nuclear Sciences, Belgrade, Serbia
\end{Authlist}
\section*{Collaboration Institutes}
\renewcommand\theenumi{\arabic{enumi}~}
\begin{Authlist}
\item \Idef{org1279}Benem\'{e}rita Universidad Aut\'{o}noma de Puebla, Puebla, Mexico
\item \Idef{org1220}Bogolyubov Institute for Theoretical Physics, Kiev, Ukraine
\item \Idef{org1262}Budker Institute for Nuclear Physics, Novosibirsk, Russia
\item \Idef{org1292}California Polytechnic State University, San Luis Obispo, California, United States
\item \Idef{org14939}Centre de Calcul de l'IN2P3, Villeurbanne, France
\item \Idef{org1197}Centro de Aplicaciones Tecnol\'{o}gicas y Desarrollo Nuclear (CEADEN), Havana, Cuba
\item \Idef{org1242}Centro de Investigaciones Energ\'{e}ticas Medioambientales y Tecnol\'{o}gicas (CIEMAT), Madrid, Spain
\item \Idef{org1244}Centro de Investigaci\'{o}n y de Estudios Avanzados (CINVESTAV), Mexico City and M\'{e}rida, Mexico
\item \Idef{org1335}Centro Fermi -- Centro Studi e Ricerche e Museo Storico della Fisica ``Enrico Fermi'', Rome, Italy
\item \Idef{org17347}Chicago State University, Chicago, United States
\item \Idef{org1288}Commissariat \`{a} l'Energie Atomique, IRFU, Saclay, France
\item \Idef{org1294}Departamento de F\'{\i}sica de Part\'{\i}culas and IGFAE, Universidad de Santiago de Compostela, Santiago de Compostela, Spain
\item \Idef{org1106}Department of Physics Aligarh Muslim University, Aligarh, India
\item \Idef{org1121}Department of Physics and Technology, University of Bergen, Bergen, Norway
\item \Idef{org1162}Department of Physics, Ohio State University, Columbus, Ohio, United States
\item \Idef{org1300}Department of Physics, Sejong University, Seoul, South Korea
\item \Idef{org1268}Department of Physics, University of Oslo, Oslo, Norway
\item \Idef{org1145}Dipartimento di Fisica dell'Universit\`{a} and Sezione INFN, Cagliari, Italy
\item \Idef{org1270}Dipartimento di Fisica dell'Universit\`{a} and Sezione INFN, Padova, Italy
\item \Idef{org1315}Dipartimento di Fisica dell'Universit\`{a} and Sezione INFN, Trieste, Italy
\item \Idef{org1132}Dipartimento di Fisica dell'Universit\`{a} and Sezione INFN, Bologna, Italy
\item \Idef{org1285}Dipartimento di Fisica dell'Universit\`{a} `La Sapienza' and Sezione INFN, Rome, Italy
\item \Idef{org1154}Dipartimento di Fisica e Astronomia dell'Universit\`{a} and Sezione INFN, Catania, Italy
\item \Idef{org1290}Dipartimento di Fisica `E.R.~Caianiello' dell'Universit\`{a} and Gruppo Collegato INFN, Salerno, Italy
\item \Idef{org1312}Dipartimento di Fisica Sperimentale dell'Universit\`{a} and Sezione INFN, Turin, Italy
\item \Idef{org1103}Dipartimento di Scienze e Tecnologie Avanzate dell'Universit\`{a} del Piemonte Orientale and Gruppo Collegato INFN, Alessandria, Italy
\item \Idef{org1114}Dipartimento Interateneo di Fisica `M.~Merlin' and Sezione INFN, Bari, Italy
\item \Idef{org1237}Division of Experimental High Energy Physics, University of Lund, Lund, Sweden
\item \Idef{org1192}European Organization for Nuclear Research (CERN), Geneva, Switzerland
\item \Idef{org1227}Fachhochschule K\"{o}ln, K\"{o}ln, Germany
\item \Idef{org1122}Faculty of Engineering, Bergen University College, Bergen, Norway
\item \Idef{org1136}Faculty of Mathematics, Physics and Informatics, Comenius University, Bratislava, Slovakia
\item \Idef{org1274}Faculty of Nuclear Sciences and Physical Engineering, Czech Technical University in Prague, Prague, Czech Republic
\item \Idef{org1229}Faculty of Science, P.J.~\v{S}af\'{a}rik University, Ko\v{s}ice, Slovakia
\item \Idef{org1184}Frankfurt Institute for Advanced Studies, Johann Wolfgang Goethe-Universit\"{a}t Frankfurt, Frankfurt, Germany
\item \Idef{org1215}Gangneung-Wonju National University, Gangneung, South Korea
\item \Idef{org1212}Helsinki Institute of Physics (HIP) and University of Jyv\"{a}skyl\"{a}, Jyv\"{a}skyl\"{a}, Finland
\item \Idef{org1203}Hiroshima University, Hiroshima, Japan
\item \Idef{org1329}Hua-Zhong Normal University, Wuhan, China
\item \Idef{org1254}Indian Institute of Technology, Mumbai, India
\item \Idef{org36378}Indian Institute of Technology Indore (IIT), Indore, India
\item \Idef{org1266}Institut de Physique Nucl\'{e}aire d'Orsay (IPNO), Universit\'{e} Paris-Sud, CNRS-IN2P3, Orsay, France
\item \Idef{org1277}Institute for High Energy Physics, Protvino, Russia
\item \Idef{org1249}Institute for Nuclear Research, Academy of Sciences, Moscow, Russia
\item \Idef{org1320}Nikhef, National Institute for Subatomic Physics and Institute for Subatomic Physics of Utrecht University, Utrecht, Netherlands
\item \Idef{org1250}Institute for Theoretical and Experimental Physics, Moscow, Russia
\item \Idef{org1230}Institute of Experimental Physics, Slovak Academy of Sciences, Ko\v{s}ice, Slovakia
\item \Idef{org1127}Institute of Physics, Bhubaneswar, India
\item \Idef{org1275}Institute of Physics, Academy of Sciences of the Czech Republic, Prague, Czech Republic
\item \Idef{org1139}Institute of Space Sciences (ISS), Bucharest, Romania
\item \Idef{org27399}Institut f\"{u}r Informatik, Johann Wolfgang Goethe-Universit\"{a}t Frankfurt, Frankfurt, Germany
\item \Idef{org1185}Institut f\"{u}r Kernphysik, Johann Wolfgang Goethe-Universit\"{a}t Frankfurt, Frankfurt, Germany
\item \Idef{org1177}Institut f\"{u}r Kernphysik, Technische Universit\"{a}t Darmstadt, Darmstadt, Germany
\item \Idef{org1256}Institut f\"{u}r Kernphysik, Westf\"{a}lische Wilhelms-Universit\"{a}t M\"{u}nster, M\"{u}nster, Germany
\item \Idef{org1246}Instituto de Ciencias Nucleares, Universidad Nacional Aut\'{o}noma de M\'{e}xico, Mexico City, Mexico
\item \Idef{org1247}Instituto de F\'{\i}sica, Universidad Nacional Aut\'{o}noma de M\'{e}xico, Mexico City, Mexico
\item \Idef{org23333}Institut of Theoretical Physics, University of Wroclaw
\item \Idef{org1308}Institut Pluridisciplinaire Hubert Curien (IPHC), Universit\'{e} de Strasbourg, CNRS-IN2P3, Strasbourg, France
\item \Idef{org1182}Joint Institute for Nuclear Research (JINR), Dubna, Russia
\item \Idef{org1143}KFKI Research Institute for Particle and Nuclear Physics, Hungarian Academy of Sciences, Budapest, Hungary
\item \Idef{org1199}Kirchhoff-Institut f\"{u}r Physik, Ruprecht-Karls-Universit\"{a}t Heidelberg, Heidelberg, Germany
\item \Idef{org20954}Korea Institute of Science and Technology Information
\item \Idef{org1160}Laboratoire de Physique Corpusculaire (LPC), Clermont Universit\'{e}, Universit\'{e} Blaise Pascal, CNRS--IN2P3, Clermont-Ferrand, France
\item \Idef{org1194}Laboratoire de Physique Subatomique et de Cosmologie (LPSC), Universit\'{e} Joseph Fourier, CNRS-IN2P3, Institut Polytechnique de Grenoble, Grenoble, France
\item \Idef{org1187}Laboratori Nazionali di Frascati, INFN, Frascati, Italy
\item \Idef{org1232}Laboratori Nazionali di Legnaro, INFN, Legnaro, Italy
\item \Idef{org1125}Lawrence Berkeley National Laboratory, Berkeley, California, United States
\item \Idef{org1234}Lawrence Livermore National Laboratory, Livermore, California, United States
\item \Idef{org1251}Moscow Engineering Physics Institute, Moscow, Russia
\item \Idef{org1140}National Institute for Physics and Nuclear Engineering, Bucharest, Romania
\item \Idef{org1165}Niels Bohr Institute, University of Copenhagen, Copenhagen, Denmark
\item \Idef{org1109}Nikhef, National Institute for Subatomic Physics, Amsterdam, Netherlands
\item \Idef{org1283}Nuclear Physics Institute, Academy of Sciences of the Czech Republic, \v{R}e\v{z} u Prahy, Czech Republic
\item \Idef{org1264}Oak Ridge National Laboratory, Oak Ridge, Tennessee, United States
\item \Idef{org1189}Petersburg Nuclear Physics Institute, Gatchina, Russia
\item \Idef{org1170}Physics Department, Creighton University, Omaha, Nebraska, United States
\item \Idef{org1157}Physics Department, Panjab University, Chandigarh, India
\item \Idef{org1112}Physics Department, University of Athens, Athens, Greece
\item \Idef{org1152}Physics Department, University of Cape Town, iThemba LABS, Cape Town, South Africa
\item \Idef{org1209}Physics Department, University of Jammu, Jammu, India
\item \Idef{org1207}Physics Department, University of Rajasthan, Jaipur, India
\item \Idef{org1200}Physikalisches Institut, Ruprecht-Karls-Universit\"{a}t Heidelberg, Heidelberg, Germany
\item \Idef{org1325}Purdue University, West Lafayette, Indiana, United States
\item \Idef{org1281}Pusan National University, Pusan, South Korea
\item \Idef{org1176}Research Division and ExtreMe Matter Institute EMMI, GSI Helmholtzzentrum f\"ur Schwerionenforschung, Darmstadt, Germany
\item \Idef{org1334}Rudjer Bo\v{s}kovi\'{c} Institute, Zagreb, Croatia
\item \Idef{org1298}Russian Federal Nuclear Center (VNIIEF), Sarov, Russia
\item \Idef{org1252}Russian Research Centre Kurchatov Institute, Moscow, Russia
\item \Idef{org1224}Saha Institute of Nuclear Physics, Kolkata, India
\item \Idef{org1130}School of Physics and Astronomy, University of Birmingham, Birmingham, United Kingdom
\item \Idef{org1338}Secci\'{o}n F\'{\i}sica, Departamento de Ciencias, Pontificia Universidad Cat\'{o}lica del Per\'{u}, Lima, Peru
\item \Idef{org1316}Sezione INFN, Trieste, Italy
\item \Idef{org1271}Sezione INFN, Padova, Italy
\item \Idef{org1313}Sezione INFN, Turin, Italy
\item \Idef{org1286}Sezione INFN, Rome, Italy
\item \Idef{org1146}Sezione INFN, Cagliari, Italy
\item \Idef{org1133}Sezione INFN, Bologna, Italy
\item \Idef{org1115}Sezione INFN, Bari, Italy
\item \Idef{org1155}Sezione INFN, Catania, Italy
\item \Idef{org1322}Soltan Institute for Nuclear Studies, Warsaw, Poland
\item \Idef{org1258}SUBATECH, Ecole des Mines de Nantes, Universit\'{e} de Nantes, CNRS-IN2P3, Nantes, France
\item \Idef{org1304}Technical University of Split FESB, Split, Croatia
\item \Idef{org1168}The Henryk Niewodniczanski Institute of Nuclear Physics, Polish Academy of Sciences, Cracow, Poland
\item \Idef{org17361}The University of Texas at Austin, Physics Department, Austin, TX, United States
\item \Idef{org1173}Universidad Aut\'{o}noma de Sinaloa, Culiac\'{a}n, Mexico
\item \Idef{org1296}Universidade de S\~{a}o Paulo (USP), S\~{a}o Paulo, Brazil
\item \Idef{org1149}Universidade Estadual de Campinas (UNICAMP), Campinas, Brazil
\item \Idef{org1239}Universit\'{e} de Lyon, Universit\'{e} Lyon 1, CNRS/IN2P3, IPN-Lyon, Villeurbanne, France
\item \Idef{org1205}University of Houston, Houston, Texas, United States
\item \Idef{org1222}University of Tennessee, Knoxville, Tennessee, United States
\item \Idef{org1310}University of Tokyo, Tokyo, Japan
\item \Idef{org1318}University of Tsukuba, Tsukuba, Japan
\item \Idef{org21360}Eberhard Karls Universit\"{a}t T\"{u}bingen, T\"{u}bingen, Germany
\item \Idef{org1225}Variable Energy Cyclotron Centre, Kolkata, India
\item \Idef{org1306}V.~Fock Institute for Physics, St. Petersburg State University, St. Petersburg, Russia
\item \Idef{org1323}Warsaw University of Technology, Warsaw, Poland
\item \Idef{org1179}Wayne State University, Detroit, Michigan, United States
\item \Idef{org1260}Yale University, New Haven, Connecticut, United States
\item \Idef{org1332}Yerevan Physics Institute, Yerevan, Armenia
\item \Idef{org15649}Yildiz Technical University, Istanbul, Turkey
\item \Idef{org1301}Yonsei University, Seoul, South Korea
\item \Idef{org1327}Zentrum f\"{u}r Technologietransfer und Telekommunikation (ZTT), Fachhochschule Worms, Worms, Germany
\end{Authlist}
\endgroup

%
%
\end{document}